\documentclass{article}

\usepackage[preprint]{neurips_2025}

\usepackage[utf8]{inputenc} 
\usepackage[T1]{fontenc}    
\usepackage{hyperref}       
\usepackage{url}            
\usepackage{booktabs}       
\usepackage{amsfonts}       
\usepackage{nicefrac}       
\usepackage{microtype}      
\usepackage{xcolor}         

\usepackage{booktabs} 
\usepackage{multirow} 
\definecolor{myblue}{RGB}{0,0,255}

\usepackage{amsmath}
\usepackage{amssymb}
\usepackage{mathtools}
\usepackage{amsthm}
\usepackage{bbding}

\usepackage{CJKutf8} 
\usepackage{microtype}
\usepackage{graphicx}
\usepackage{subfigure}
\usepackage{subcaption}

\usepackage{arydshln} 
\usepackage{wrapfig}  

\usepackage{threeparttable}
\usepackage{makecell}
\usepackage{siunitx}
\usepackage{textcomp}       

\title{U-Codec: Ultra Low Frame-rate Neural Speech Codec for Fast High-fidelity Speech Generation}

\author{%
  Xusheng Yang$^{\star}$,
    Long Zhou$^{\diamond}$, Wenfu Wang$^{\spadesuit}$, Kai Hu$^{\diamond}$, 
    Shulin Feng$^{\diamond}$, Chenxing Li$^{\spadesuit}$, \\
    \textbf{Meng Yu$^{\spadesuit}$, Dong Yu$^{\spadesuit}$, 
    Yuexian Zou$^{\star\dagger}$}\thanks{
    $\dagger$ Corresponding author} \\
    $^{\star}$ Peking University, 
    $^{\spadesuit}$ Tencent AI Lab,
    $^{\diamond}$ Tencent Hunyuan\\
    \texttt{2201111705@pku.stu.edu.cn}
}

\begin{document}

\maketitle

\begin{CJK*}{UTF8}{gbsn}
\begin{abstract}

We propose \textbf{U-Codec}, an \textbf{U}ltra low frame-rate neural speech \textbf{Codec} that achieves high-fidelity reconstruction and fast speech generation at an extremely low frame-rate  of 5Hz (5 frames per second). 
Extreme compression at 5Hz typically leads to severe intelligibility and spectral detail loss, we introduce a Transformer-based inter-frame long-term dependency module and systematically explore residual vector quantization (RVQ) depth and codebook size to identify optimal configurations.
Moreover, we apply U-Codec into a large language model (LLM)-based auto-regressive TTS model, which leverages global and local hierarchical architecture to effectively capture dependencies across multi-layer tokens. 
We extend LLM-based TTS from 3-layer RVQ at 50Hz to 32-layer RVQ at 5Hz.
Experimental results demonstrate that U-Codec improves LLM-based TTS inference speed by around 3 $\times$ over high-frame-rate codecs while maintaining similarity and naturalness. 
These results validate the feasibility of using highly compressed 5Hz discrete tokens for fast and high-fidelity speech synthesis.





\end{abstract}
\end{CJK*}

\begin{figure}[hbt]
\centering
\centerline{\includegraphics[width=0.8\textwidth]{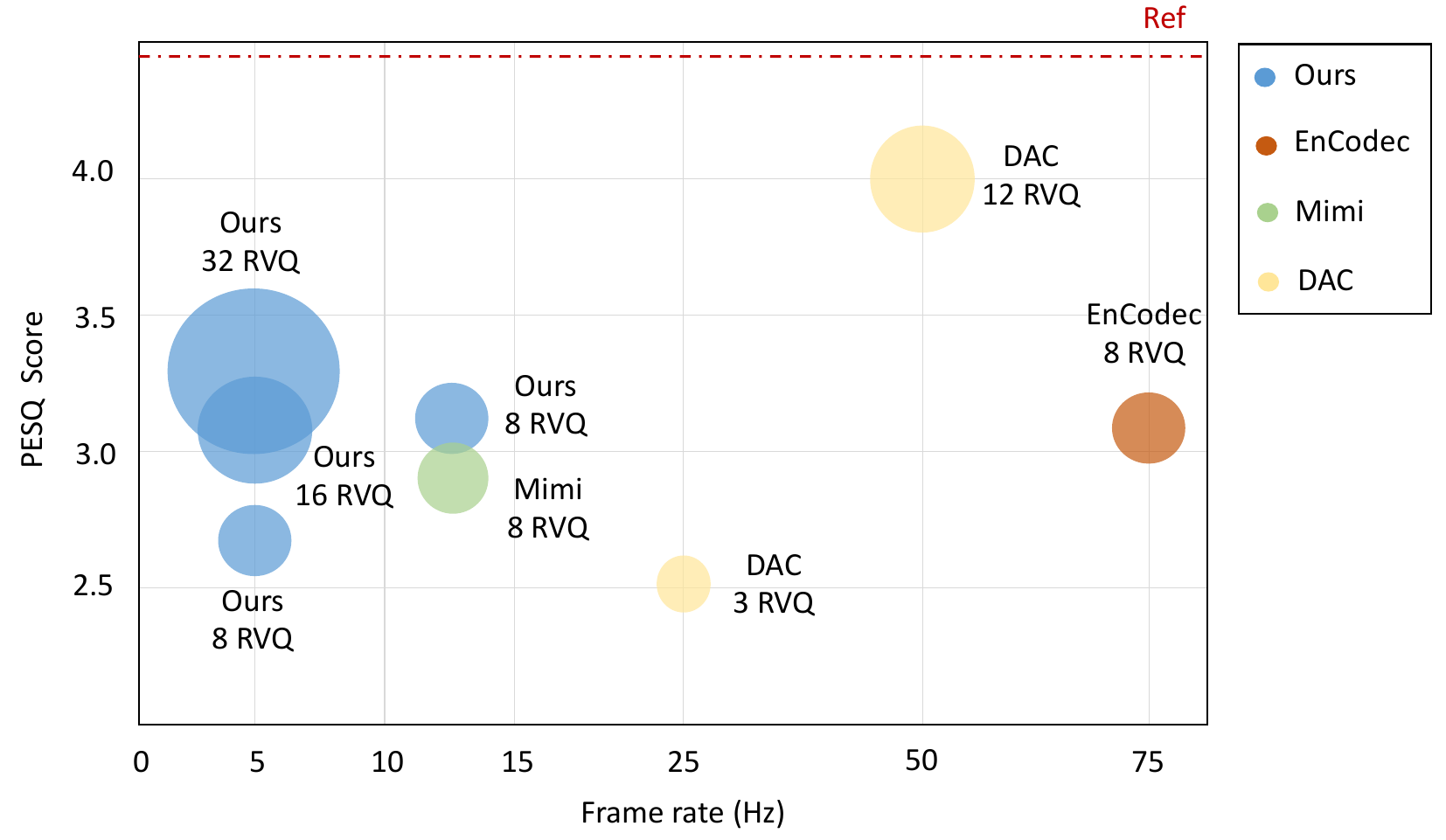}}
\caption{Reconstruct quality on varying the frame rate of different codecs, where bubble size indicates the number of RVQ layers. 
Compared to previous systems, such as EnCodec, Mimi, and DAC operating at higher frame rates, our proposed codec achieves competitive PESQ performance under the ultra low frame rate (5Hz).
}
\label{fig:1}
\end{figure}

\newpage

\section{Introduction}
\label{section1}

Neural speech codecs (NSCs) are essential signal processing techniques that compress continuous speech signals into discrete codes and reconstruct the original signal from these codes. 
Traditionally, NSCs have been widely used in communication and transmission tasks \cite{zeghidour2021soundstream, defossez2022high, ahn2024hilcodec, yangvchangecodec}.
Recently, NSCs \cite{pons2021upsampling, yang2023hifi,wu2023audiodec} have driven significant progress in zero-shot text-to-speech synthesis (TTS) \cite{wang2023neural, zhang2023speak, wang2024speechx, chen2024vall, borsos2023soundstorm, ye2025llasa}, particularly as tokenizers that convert continuous speech signals into discrete tokens for large language models (LLM)-based speech generation systems and de-tokenizers that reconstruct speech signals from discrete tokens. 
Additionally, these discrete speech tokens allow LLMs to process acoustic and textual information within a unified framework, establishing the foundation for cross-modal alignment and generation.

Current NSCs target low bitrates \cite{ai2024low, ji2024wavtokenizer, wan2025spectokenizer} with high fidelity via residual vector quantization (RVQ) and advanced loss designs. However, state-of-the-art (SOTA) systems such as SoundStream, EnCodec, DAC, and APCodec operate at high frame rates with 50-75 frames per second (FPS), which is inefficient for auto-regressive LLM-based TTS since each frame requires a forward pass.
In auto-regressive TTS systems, each frame requires one forward pass, meaning high frame rates lead to long token sequences and slow inference. 
For example, the DAC codec produces audio at 75 FPS, and TTS needs to make at least 75 forward passes to predict one second of audio.
As a result, high frame rates substantially increase inference speed and computational complexity.

Reducing the frame rate provides a direct way to shorten token sequences and accelerate LLM-based TTS. 
Prior work \cite{chen2024vall, han2024vall} has shown that reducing the frame rate is essential, but not straightforward. 
Recent studies on speech tokenizers \cite{defossez2024moshi, yang2025almtokenizer, li2025dualcodec} with a moderate frame rate of 12.5Hz report a sharp drop in reconstruction quality (e.g., PESQ around 2.0) and incorrect phoneme pronunciation due to loss of fine-grained acoustic details.
Moreover, pushing the frame rate to a more extreme level, such as 5Hz, has not been systematically investigated.
This inspires us to ask one question: \textbf{Can discrete speech tokens at 5Hz achieve both high-quality reconstruction and high-efficient speech synthesis?}
To explore this, we have conducted preliminary experiments varying the frame rate and RVQ depth across different codecs. 
As illustrated in Figure \ref{fig:1} (a), our proposed codec achieves competitive PESQ scores at ultra-low frame rates (5–12.5Hz), where existing systems such as EnCodec, Mimi \cite{defossez2024moshi}, and DAC require much higher frame rates to maintain performance, indicating that multi-layer RVQ with long-term dependency modeling can enable first 5Hz speech tokenizer for LLM-based TTS.



In this paper, we propose the U-Codec, an ultra-low frame rate neural speech codec that that achieves high-fidelity speech compression at 5Hz, significantly improving both computational efficiency and inference speed of LLM-based TTS. 
To address loss of speech and spectrum details at extreme compression, we introduce a Transformer-based inter-frame long-term dependency model that improves fidelity and carefully optimize residual vector quantization (RVQ) layers and codebook sizes. 
However, while increasing the number of RVQ layers improves speech quality, it also raises TTS inference cost, as more codes must be predicted per frame.
Inspired by long discrete audio sequences modeling \cite{yu2023megabyte, yang2023uniaudio}, we introduce a Codecformer network which is a hierarchical model that processes the inter- and intra-frame correlation of audio separately. 
For example, with an 8-layer RVQ at 5Hz, only 5 forward passes are used for the global network while all 8 are handled for the local network. 
This design allows for efficient joint modeling of discrete tokens at multiple layers (e.g., 16 or 32 layers) for downstream TTS tasks. 
Experimental results show that U-Codec achieves a about 3 $\times$ speedup in inference compared to existing high-frame-rate codecs, establishing the feasibility of discrete speech tokens at 5Hz for fast, high-fidelity speech generation.

Our contributions are summarized as follows: 

\begin{itemize}

\item We present \textbf{U-Codec}, the first ultra low frame-rate speech codec that operates at 5Hz with extreme compression.
This provides a fully discrete speech tokenizer for efficient speech–text cross-modal modeling in LLM-based systems.

\item  We systematically optimize residual vector quantization (RVQ) depth and codebook size, and introduce a Transformer-based inter-frame dependency module to preserve intelligibility and spectral detail under extreme compression.  

\item We introduce Codecformer network with global and local hierarchical architecture, reduce the cost of TTS inference while maintaining scalability to deeper RVQ stacks.

\item The comprehensive evaluations show that U-Codec improves LLM-based TTS inference speed by about 3 $\times$ compared with high-frame-rate codecs while maintaining SOTA quality, and \bf we release the demo and code to the community \footnote{\href{https://yangxusheng-yxs.github.io/U-Codec/}{https://yangxusheng-yxs.github.io/U-Codec/}}.

\end{itemize}

\section{Related Work}
\label{related work}

\paragraph{Neural audio codec models.}

Neural audio codecs (NACs) discretize continuous speech signals into tokens, supporting both efficient compression and large language models (LLM)-based speech generation. 
Early systems such as SoundStream, EnCodec and DAC \citep{zeghidour2021soundstream, defossez2022high, kumar2024high} achieve impressive reconstruction quality but typically operate at high frame rates (e.g., more than 50Hz),  which limits modeling efficiency in LLM-based frameworks. 
To utilize the LLM ability on speech data, subsequent studies have focused on reducing bitrate while enriching the semantic representation of audio tokens \cite{ye2025llasa, ye2025codec, zhang2024speechtokenizer, zhang2025vevo, liu2024semanticodec}. In parallel, several works \cite{defossez2024moshi, li2025dualcodec} explored lower frame rates to improve efficiency. 
For example, by utilizing Transformer models and SpeechTokenizer, Mimi \cite{defossez2024moshi} demonstrated that a NAC model can achieve a frame rate as low as 12.5Hz.
DualCodec \cite{li2025dualcodec} introduces a dual-stream encoding approach, combining self-supervised learning (SSL), achieving high-fidelity and semantically enhanced speech reconstruction at 12.5 Hz.
ALMTokenizer \cite{yang2025almtokenizer} further proposes a query-based compression strategy that yields semantically rich representations at the same frame rate.
Despite these advancements, achieving ultra-low frame rates (e.g., 5Hz) with high reconstruction quality remains an unsolved problem. 
In this work, we present a codec that maintains high fidelity at an unprecedented 5Hz frame rate, which is about 2.5 $\times$ lower than Mimi, explicitly exploring the trade-off between multi-layer RVQ depth and codebook size.

\paragraph{Zero-shot TTS.}

Recent zero-shot text-to-speech (TTS) systems \cite{wang2023neural, anastassiou2024seed, du2024cosyvoice, wang2025spark, peng2024voicecraft} leverage discrete speech tokens to synthesize voices for unseen speakers using short reference audio.
Early methods \cite{casanova2022yourtts, chen2018sample} relied on explicit speaker embeddings or adaptation strategies, while recent studies integrate large language models (LLMs) with neural audio codecs (NACs).
VALL-E \cite{chen2024vall} pioneered the LLM-based TTS paradigm, inspiring extensions including VALL-E X \cite{zhang2023speak} for cross-lingual synthesis and VALL-E 2 \cite{chen2024vall} with grouped code modeling, and variants incorporating prosody features and alignment mechanisms. 
While these approaches demonstrate impressive zero-shot capability, they remain computationally demanding due to auto-regressive generation over long token sequences.
Parallel to autoregressive designs, diffusion-based frameworks such as NaturalSpeech and Voicebox \cite{tan2024naturalspeech, le2023voicebox}, and prompt-driven systems like Prompt-TTS \cite{leng2023prompttts}, achieve natural and controllable speech generation. 
Beyond TTS, some studies have also proposed audio generation models, such as UniAudio and UniAudio 1.5 \cite{yang2023uniaudio, yang2024uniaudio}, which use hierarchical modeling to reduce the sequence length of tokens processed by Transformer.
Our work directly introduces the first 5Hz speech tokenizer optimized for LLM-based TTS, demonstrating that ultra low frame-rate tokens can preserve reconstruction quality while accelerating TTS inference.

\section{Methodology}
\label{U-Codec}
\begin{figure*}[ht]
\centering
\centerline{\includegraphics[width=0.8\textwidth]{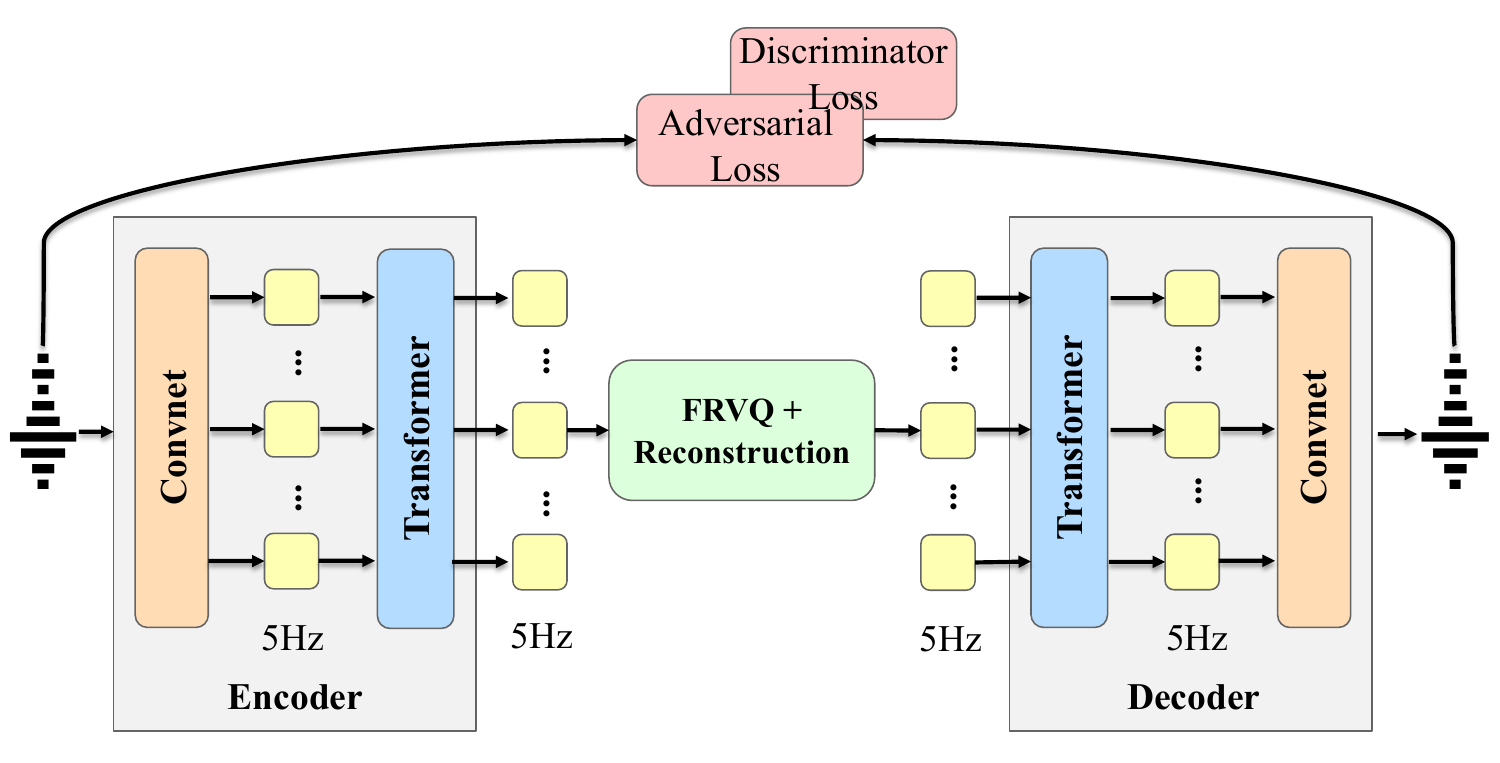}}
\caption{Architecture and training of our neural speech codec at 5Hz. 
The encoder (left) consists of convolutional layers followed by a Transformer to capture long-term dependencies. 
Latent features are quantized through residual vector quantization (RVQ) and optimized for high-fidelity reconstruction. 
The decoder (right) mirrors the encoder to synthesize the output waveform. 
We use reconstruction loss and adversarial loss during training.}
\label{fig:2}
\end{figure*}

A schematic diagram of U-Codec is shown in Fig. \ref{fig:2}. 
Referring to the existing neural speech codec such as SoundStream, Encodec, DAC and MimiCodec \cite{zeghidour2021soundstream, defossez2022high, kumar2024high, defossez2024moshi},
we meticulously optimized every component of U-Codec, including the encoder, decoder, quantizer, and a HiFi-GAN-based discriminator, to achieve high-quality speech compression under ultra low frame rates.

\subsection{Structure of U-Codec}
\label{section3.1}
\textbf{Encoder.}
The encoder transforms a 16kHz waveform into compact latent representations through five residual convolutional blocks with dilated and strided convolutions along with ELU non-linearities and weight normalization. 
To achieve 5Hz (i.e., one token per 200 ms) temporal resolution, each residual block is followed by a strided 1D convolution layer with strides of $[8, 5, 5, 4, 4]$. 
The encoder starts with 64 channels and doubles the width after the final downsampling stage, progressively compressing the temporal resolution while increasing feature capacity.

\textbf{Long-term Dependency Module.}
At an ultra low frame-rate of 5Hz, each token covers a long speech span, which may degrade intelligibility and spectral fidelity.
To address this, we introduce a contextual Transformer bottleneck directly after downsampling.
The Transformer consists of 8 layers, each with 8 attention heads, RoPE position encodings, GELU activations, hidden size 512, and MLP dimension 2048. 
Unlike MimiCodec, which applies convolutional compression to 12.5Hz after the Transformer module, U-Codec performs all temporal compression before the Transformer. 
This design explicitly models inter-frame dependencies on the ultra low frame-rate sequence, thereby preserving accurate phonetic alignment and spectral continuity.

\textbf{Decoder.} 
The decoder mirrors the encoder’s downsampling structure with upsampling factors of $[4, 4, 5, 5, 8]$. The decoder begins with 2048 channels and halves the channel width at each stage, synthesizing high-fidelity 16kHz speech.

\textbf{Frame rate.} 
The 5Hz frame rate results from the cumulative stride $(8×5×5×4×4)$ applied in the encoder, corresponding to $16000 \div 3200 = 5$Hz.
The 12.5Hz version uses strides $(5, 4, 4, 4, 4)$ for higher temporal resolution.

\textbf{RVQ Module and Quantization Rate.}  
To improve codebook utilization and achieve stable reconstruction, we adopt factorized residual vector quantization (FRVQ). FRVQ combines two key techniques: (i) \textit{factorized coding}, where code lookup is performed in a compact low-dimensional space (e.g., 8 dimensions) while code embedding is maintained in a high-dimensional space  (e.g., 1024 dimensions), enabling the quantizer to capture the principal variations of the input with reduced redundancy; and (ii) \textit{L2-regularized coding}, which replaces Euclidean distance with cosine similarity, improving stability and reconstruction fidelity. 
Compared with a vanilla RVQ quantizer, this design results in more efficient codebook utilization and higher-quality speech compression.  

Building on this foundation, we systematically explore RVQ under extreme compression conditions (5Hz), where the bitrate remains around 1kbps. 
Guided by the entropy formulation (the resulting bitrate) $ S\times N \times \log_2 C$, $S$ is the frame size, $N$ the number of quantization layers and $C$ denotes codebook size, we explore a wide range of configurations from shallow (8 layers, large codebooks up to 16,384) to deep (100 layers, small codebooks down to 4). 
This comprehensive study provides valuable insights into the trade-off between reconstruction quality, token compactness, and bitrate constraints. 
Finally, our design balances sequence length reduction with sufficient acoustic detail, enabling U-Codec to operate effectively at 5Hz for high-quality downstream TTS.

\textbf{Discriminator.} To stabilize adversarial training, we adopt discriminators from BigCodec \cite{xin2024bigcodec}. 
Specifically, we employ the HiFiGAN Multi-Period Discriminator (MPD) \cite{kong2020hifi} and the multi-scale short-time Fourier transform (MS-STFT) with FFT sizes $\{78, 126, 206, 334, 542, 876, 1418, 2296\}$ \cite{ye2025llasa, parker2024scaling}. 
Preliminary experiments confirmed that these two components effectively improve perceptual quality, while adding a Multi-Scale Discriminator (MSD) did not provide additional benefit.

\subsection{Training Strategy}

The proposed U-Codec is trained with a weighted combination of reconstruction, adversarial, feature matching, and vector quantization losses. The design of the training objective is motivated by prior studies on neural speech codecs such as DAC~\cite{kumar2024high}, HiFi-GAN~\cite{kong2020hifi}, and BigCodec~\cite{xin2024bigcodec}, while adapted to the ultra low frame-rate and multi-layer FRVQ setting of our model.
\textbf{Reconstruction loss.}  
Following DAC, we adopt the multi-scale mel-spectrogram loss, which measures the $L_1$ distance between predicted and ground-truth log-mel features at multiple resolutions.
Compared to multi-scale STFT losses, this design correlates more closely with perceptual quality.
\textbf{Adversarial loss.}  
We employ least-squares GAN (LSGAN) \cite{mao2017least} with a multi-period discriminator~\cite{kong2020hifi}. 
The adversarial loss encourages generated waveforms $\hat{x}$ to lie close to the real data manifold to stabilize training. 
\textbf{Feature matching loss.}  
We apply the $L_1$ feature matching loss~\cite{kong2020hifi} between intermediate discriminator activations of real and generated speech.
\textbf{Vector quantization loss.}  
The codebook is optimized using an  $L_1$  loss between the projected encoder output and the quantized result, with a stop-gradient operator applied. 
To prevent encoder outputs from growing excessively, a commitment loss is introduced. 
The final loss is the weighted sum of all terms.
The multi-scale mel-spectrogram loss is given a weight of 15 due to its strong impact on speech quality. The VQ commitment loss is weighted at 0.25 to avoid mode collapse, while all other losses use a weight of 1.


\begin{figure*}[ht]
\centering
\centerline{\includegraphics[width=0.8\textwidth]{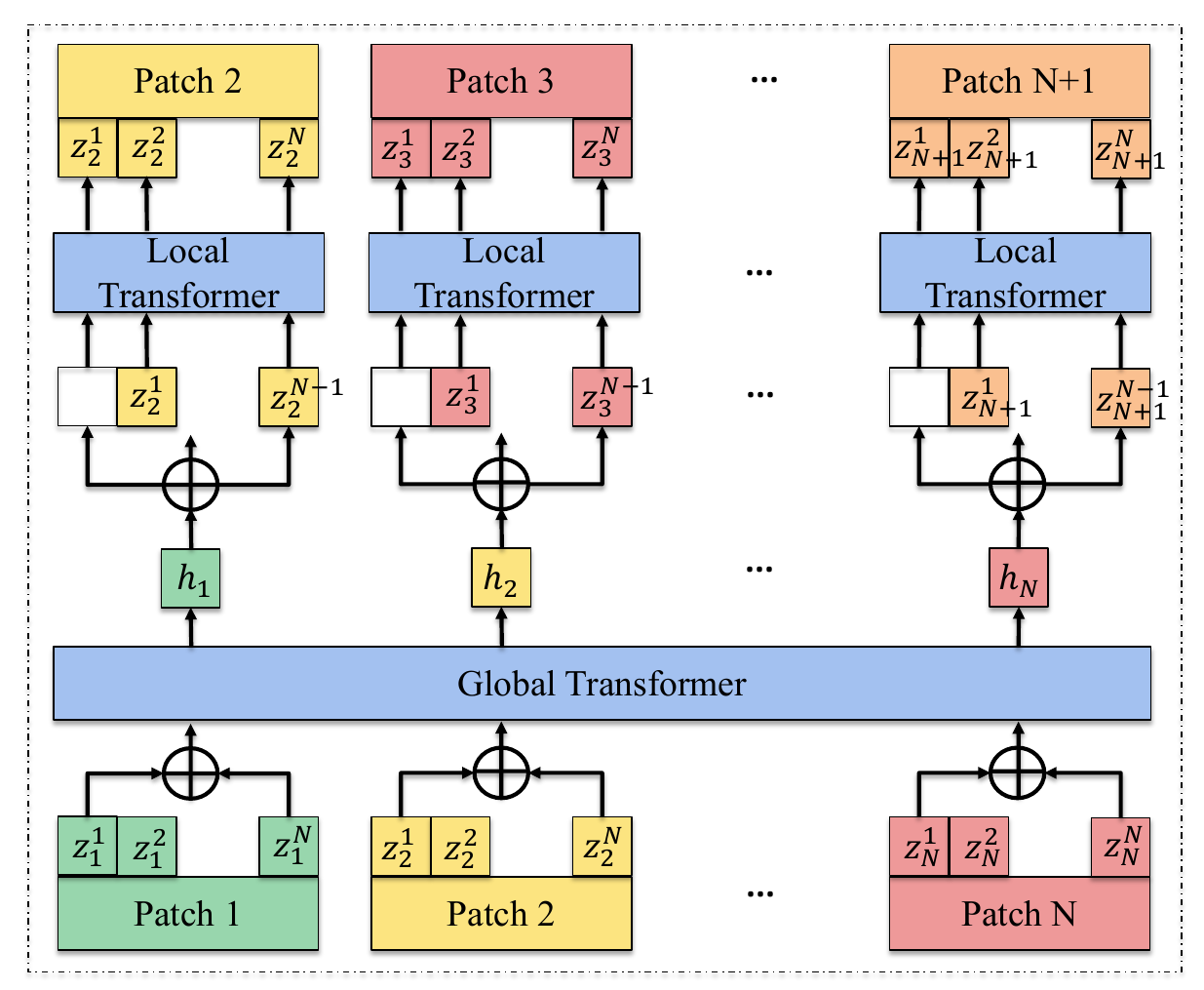}}
\caption{Hierarchical global-local Transformer architecture. Each input sequence is divided into patches, where $z^{k}_{N}$ denotes the $k-th$ token in the $t-th$ patch.}
\label{fig:3}
\end{figure*}

\subsection{Speech Language Modeling with U-Codec}
\label{section3.2}

A major challenge in speech language modeling is the scalability of autoregressive Transformers when applied to long token sequences. Flattened token modeling requires processing sequences of length $T \times N$, where $T$ is the number of frames and $N$ the number of quantization layers, leading to quadratic complexity in self-attention ~\cite{vaswani2017attention}.
This issue becomes especially severe for multi-layer RVQ systems at ultra-low frame rates (e.g., 5Hz), where $N$ can reach 32 or even 100.

To address this problem, we introduce a hierarchical global-local Transformer architecture inspired by CodecFormer~\cite{yang2023uniaudio, yu2023megabyte}, which explicitly decouples inter-frame and intra-frame dependencies as shown in Fig. \ref{fig:3}. Each frame’s $N$ tokens are grouped into a patch, reducing the sequence length for global modeling from $T \times N$ to $T$.  

\textbf{Global Transformer.}  
Each patch $z_t$ is aggregated by summing token embeddings, consistent with RVQ aggregation. The global Transformer then models long-range dependencies across frames:
\begin{equation}
    h_t = f_{\text{global}}(z_{\leq t}).
\end{equation}

\textbf{Local Transformer.}  
Conditioned on $h_t$, the local Transformer autoregressively predicts the tokens within the next patch:
\begin{equation}
    p(z_{t+1} \mid h_t) = \prod_{k=1}^N p(z_{t+1}^k \mid z_{t+1}^{<k}, h_t).
\end{equation}

This hierarchical design provides \textbf{three main advantages}.
First, reducing the global sequence length from $T \times N$ to $T$ ensures scalability even when $N$ increases to 100.
Second, the local Transformer captures fine-grained intra-frame structure, enabling high-fidelity reconstruction.
Third, since the global network typically requires more Transformer layers and contributes most of the computational cost, lowering the frame rate significantly reduces the global modeling overhead—yielding large efficiency gains without compromising quality.
While CodecFormer was originally validated with only 3 RVQ layers at 50Hz, we significantly extend its applicability by exploring 8, 16, 32, and 100 layers at 5Hz. This systematic study demonstrates that our design can maintain training feasibility, and significantly accelerate inference speed at ultra-low frame rates.


\section{Experiments}
\label{experiments}
\subsection{Speech Codec Experiments}
\subsubsection{Experiments Details}
\textbf{Dataset.} 
Our U-Codec is trained on a large-scale multilingual corpus containing approximately 115k hours of speech, including LibriLight (60k hours) \cite{kahn2020libri}, GigaSpeech (10k hours) \cite{chen2021gigaspeech}, and Multilingual LibriSpeech (45k hours) \cite{pratap2020mls}, as summarized in Table \ref{tab:codec-tts}.
All recordings are resampled to 16kHz.
We evaluate reconstruction performance on the LibriSpeech test-clean set, which includes 2620 utterances at 16 kHz and serves as a standard benchmark for codec assessment.

\textbf{Training Setup.} The codec model is trained for 600k steps with a learning rate of $1 \times 10^{-4} $ and a 1000-step warmup schedule. 
We use different codebook sizes and employ a projection dimension of 8 in our vector quantization (VQ) module. We employ 16 H20 GPUs with a batch size of 16 for training. 
For the proposed 5Hz frame-rate model, we experiment with variable RVQ layer depths of $\{8, 16, 32, 100\}$ to investigate the effect of quantization hierarchy under ultra low frame conditions.

\begin{table}[htbp]
\centering
\caption{Training corpora for speech codec and TTS tasks and their respective standard test sets.}
\label{tab:codec-tts}
\resizebox{0.9\textwidth}{!}{
\begin{tabular}{@{}lcccc@{}}
\toprule
Task & Training Dataset & Training Volume (hrs) & Test Set & Test Volume (hrs)\\ \midrule
\multirow{3}{*}{Speech codec}
 & LibriLight & 60\,000 & \multirow{3}{*}{LibriSpeech test-clean} & \multirow{3}{*}{8} \\
 & GigaSpeech & 10\,000 & \\
 & MLS   & 45\,000 & \\ \midrule
TTS & LibriHeavy & 50\,000 & LibriSpeech test-clean subset & about 4 \\
\bottomrule
\end{tabular}
}
\end{table}

\subsubsection{Results and Discussion.}
\textbf{Reconstruction.} We evaluate U-Codec using WER \footnote{\href{https://huggingface.co/facebook/hubert-large-ls960-ft}{https://huggingface.co/facebook/hubert-large-ls960-ft}}, 
STOI, 
PESQ,
SPK-SIM 
\footnote{\href{https://github.com/microsoft/UniSpeech/tree/main/downstreams/speaker_verification}{https://github.com/microsoft/UniSpeech/tree/main/downstreams/speaker\_verification}}, 
and
UTMOS \footnote{\href{https://github.com/tarepan/SpeechMOS}{https://github.com/tarepan/SpeechMOS}} referring to X-Codec2 \cite{ye2025llasa}. Results are compared against a range of state-of-the-art codecs including DAC, SpeechTokenizer, BigCodec, EnCodec, WavTokenizer, Mimi, and SemanticCodec.
As summarized in Table~\ref{tab:main_reslut}, the U-Codec with 32 RVQ layers at 5Hz achieves WER~3.44, PESQ~3.20, and STOI~0.93, surpassing prior low frame-rate codecs and matching the intelligibility of much higher frame-rate systems such as BigCodec with 0.93 (5Hz vs 50Hz). 
Increasing RVQ depth from 16 to 32 layers improves PESQ (3.02$\rightarrow$3.20) and SPK-SIM (0.83$\rightarrow$0.87), confirming the effectiveness of deeper residual quantization when operating at 5 Hz.
Although U-Codec does not yet match the PESQ~4.15 and SPK-SIM~0.95 reported for certain high frame-rate systems such as DAC, our method narrows the gap and maintains strong trade-offs of quality and inference speed in Section \ref{complexity}, making it well-suited for LLM-based TTS where inference speed is critical.
Together, these results establish a new benchmark for ultra low frame-rate speech codings.

\begin{table*}[!t]
\centering
\small
\caption{Comparison between different codec models. Bold values indicate the best for each frame-rate range. 
We use frame rate instead of bitrate because, from the perspective of LLMs, it is more intuitive to influence the speed of inference.
}
\begin{threeparttable}
\resizebox{0.9\textwidth}{!}{
\begin{tabular}{lccccccccccc}
\toprule
\multirow{2}{*}{Model} & 
\multirow{2}{*}{\makecell{Frame\\Rate}} & 
\multirow{2}{*}{\makecell{Token\\Rate}} & 
\multirow{2}{*}{\makecell{Codebook\\Size}} & 
\multirow{2}{*}{\makecell{Codebook\\Layer}} & 
\multirow{2}{*}{\makecell{Bit\\(kbps)}} & 
\multirow{2}{*}{\makecell{WER\\↓}} & 
\multirow{2}{*}{\makecell{STOI\\↑}} & 
\multirow{2}{*}{\makecell{PESQ\\WB↑}} & 
\multirow{2}{*}{\makecell{PESQ\\NB↑}} & 
\multirow{2}{*}{\makecell{SPK-\\SIM↑}} & 
\multirow{2}{*}{UTMOS↑} \\
 & & & & & & & & & & \\
\midrule
Ground Truth  & - & - & - & - & - & 1.96 
& 1.00 & 4.64 & 4.55 & 1.00 & 4.09 \\
\midrule
BigCodec & 80 & 80 & 8192 & 1 & 1.04 & 2.76 & 0.93 & 2.68 & \bf 3.27 & 0.84 & \bf 4.11 \\
Encodec & 75 & 600 & 1024 & 8 & 6.00 & 2.15 & \bf 0.94 & \bf 2.77 & 3.18 & \bf 0.89 & 3.09 \\
WavTokenizer & 75 & 75 & 4096 & 1 & -- & \bf 3.98 & 0.90 & 2.13 & 2.63 & 0.65 & 3.79 \\
\midrule
DAC & 50 & 600 & 1024 & 12 & 6.00 & \bf 2.00 & \bf 0.95 & \bf 4.01 & \bf 4.15 & \bf 0.95 & \bf 4.00 \\
SpeechTokenizer & 50 & 100 & 1024 & 2 & 2.00 & 3.92 & 0.77 & 1.25 & 1.59 & 0.36 & 2.28 \\
X-codec & 50 & 100 & 1024 & 2 & 2.00 & 2.49 & 0.86 & 2.33 & 2.88 & 0.72 & 4.21 \\
DAC & 50 & 50 & 1024 & 1 & 0.50 & 74.55 & 0.62 & 1.06 & 1.20 & 0.08 & 1.25 \\
X-codec2 & 50 & 50 & 65536 & 1 & 0.80 & 2.47 & 0.92 & 2.43 & 3.04 & 0.82 & 4.13 \\
\midrule
StableCodec & 25 & 50 & 15625 & 2 & -- & 5.12 & 0.91 & 2.24 & 2.91 & 0.62 & \textbf{4.23} \\
SemantiCodec & 25 & 50 & 32768/8192 & 2 & 0.70 & 6.89 & 0.84 & 1.66 & 2.18 & 0.58 & 2.71 \\
Mimi & 12.5 & 100 & 2048 & 8 & 1.10 & 2.96 & 0.91 & 2.25 & 2.80 & 0.73 & 3.56 \\
DualCodec & 12.5 & 75 & 16384/4096 & 6 &  0.93 & - & 0.92 & \textbf{2.54} & 3.11 & - & 4.11 \\
\bf U-Codec (ours) & 12.5 & 100 & 1024 & 8 & 1.00 & \bf \textbf{2.96} & \bf \textbf{0.93} & 2.49 & \bf \textbf{3.15} & \bf \textbf{0.85} & 3.74 \\
\midrule
\bf U-Codec (ours) & 5 & 80 & 4096 & 16 & 0.96 & \bf \textbf{3.34} & 0.92 & 2.41 & 3.02 & 0.83 & \bf \textbf{3.51} \\
\bf U-Codec (ours) & 5 & 160 & 256 & 32 & 1.28 & 3.44 & \bf \textbf{0.93} & \bf \textbf{2.59} & \bf \textbf{3.20} & \bf \textbf{0.87} & 3.48 \\
\bottomrule
\end{tabular}
}
\end{threeparttable}
\label{tab:main_reslut}
\end{table*}

\textbf{Ablation Study.} 
We conduct ablation studies to analyze the contributions of key components. Table \ref{tab:ablations} summarizes results under varying configurations, focusing on the impact of RVQ depth, codebook size, and Transformer architecture.
The systematic increase in RVQ layers demonstrates clear performance improvements. This validates our hypothesis that deeper quantization compensates for the loss of temporal resolution at ultra-low frame rates. This demonstrates that enhanced residual modeling capacity is essential when the temporal sampling rate is drastically reduced.
Codebook size optimization reveals interesting trade-offs. 
Within the 8-layer configurations, increasing codebook size from 8192 to 16384 improves most metrics (WER: 5.41$\rightarrow$5.04, PESQ: 1.95$\rightarrow$2.07) but with diminishing returns and a slight bitrate increase (0.52$\rightarrow$0.56kbps). 
These results highlight a clear trade-off between representational capacity and compression efficiency. 
Further scaling to 100 layers provides only marginal additional benefits.
Crucially, replacing the Transformer with convolution module leads to consistent degradation (WER: 3.44$\rightarrow$5.40, PESQ: 2.59$\rightarrow$2.55, SPK-SIM: 0.87$\rightarrow$0.84). While the drops appear moderate, they reveal a structural limitation of convolution-only designs under ultra low frame rates: 
shift-invariant operations cannot adaptively reallocate modeling capacity across information-dense and sparse segments of speech. In contrast, the Transformer’s global attention dynamically emphasizes informative frames and suppresses redundant or silent regions, leading to more efficient long-range modeling.

\begin{table}[htbp]
\centering
\small
\caption{Comparison on different configurations of U-Codec.}
\begin{threeparttable}
\resizebox{0.9\textwidth}{!}{
\begin{tabular}{lccccccccccc}
\toprule
\multirow{2}{*}{Model} & 
\multirow{2}{*}{\makecell{Frame\\Rate}} & 
\multirow{2}{*}{\makecell{Token\\Rate}} & 
\multirow{2}{*}{\makecell{Codebook\\Size}} & 
\multirow{2}{*}{\makecell{Codebook\\Layer}} & 
\multirow{2}{*}{\makecell{Bit\\(kbps)}} & 
\multirow{2}{*}{\makecell{WER\\↓}} & 
\multirow{2}{*}{\makecell{STOI\\↑}} & 
\multirow{2}{*}{\makecell{PESQ\\WB↑}} & 
\multirow{2}{*}{\makecell{PESQ\\NB↑}} & 
\multirow{2}{*}{\makecell{SPK-\\SIM↑}} & 
\multirow{2}{*}{UTMOS↑} \\
 & & & & & & & & & & \\
\midrule
U-Codec  & 5 & 40  & 8192  & 8   & 0.52 & 5.41 & 0.89 & 1.95 & 2.50 & 0.68  & 3.31 \\
U-Codec  & 5 & 40  & 16384 & 8   & 0.56 & 5.04 & 0.90 & 2.07 & 2.67 & 0.72  & 3.47 \\
U-Codec  & 5 & 80  & 4096  & 16  & 0.96 & 3.34 & 0.92 & 2.41 & 3.02 & 0.83  & \bf 3.51 \\
U-Codec  & 5 & 160 & 256   & 32  & 1.28 & 3.44 & \bf \textbf{0.93} & \bf \textbf{2.59} & \bf \textbf{3.20} & \bf \textbf{0.87} & 3.48 \\
U-Codec  & 5 & 500 & 4     & 100 & 1.00 & \bf 2.94 & 0.92 & 2.32 & 2.90 & 0.81  & 3.41 \\
-w/o Transformer  & 5 & 160 & 256   & 32  & 1.28 & 5.40 & 0.92 & 2.55 & 3.13 & 0.84 & 3.51 \\
\bottomrule
\end{tabular}
}
\end{threeparttable}
\label{tab:ablations}
\end{table}

\begin{table}[htbp]
\centering
\caption{Comparison with prior SOTA TTS systems. Following VALL-E \cite{wang2023neural}. SIM-o measures similarity to ground-truth speech, while SIM-r measures similarity to reconstructed speech.}
\label{tab:tts_objective}
\small
\resizebox{0.9\textwidth}{!}{
\begin{tabular}{lccccc}
\toprule
\textbf{TTS Model} & \textbf{Frame Rate (Hz)} & \textbf{Zero-shot} & \textbf{SIM-r ($\uparrow$)} & \textbf{SIM-o ($\uparrow$)} & \textbf{WER ($\downarrow$)} \\
\midrule
GroundTruth & - & - & - & - & 1.9 \\
\midrule
YourTTS \cite{casanova2022yourtts} & - & \checkmark & 0.337 & 0.31 & 7.7 \\
VALL-E \cite{wang2023neural} & 75 & \checkmark & 0.580 & - & 5.9 \\
Make-A-Voice (TTS) \cite{huang2023make} & 50 & \checkmark & 0.498 & 0.45 & 5.7 \\
NaturalSpeech 2 \cite{shen2024naturalspeech} & - & \checkmark & 0.62 & 0.55 & 2.3 \\
SPEAR-TTS \cite{kharitonov2023speak} & 50 & \checkmark & 0.560 & - & - \\
VoiceBox \cite{le2023voicebox} & 100 & \checkmark & \textbf{0.681} & \textbf{0.66} & \textbf{1.9} \\
UniAudio \cite{yang2024uniaudio} & 50 & \checkmark & 0.64 & - & 2.4 \\
CLaM-TTS \cite{kimclam} & - & \checkmark & 0.5382 & 0.4951 & 5.11 \\
DiTTo-TTS \cite{lee2024ditto} & - & \checkmark & 0.6752 & 0.6538 & 2.64 \\
\midrule
\textbf{U-Codec-8RVQ-c1024} & 12.5 & \checkmark & \textbf{0.71} & \textbf{0.642} & 2.16 \\
\textbf{U-Codec-8RVQ-c16384} & 5 & \checkmark & 0.62 & 0.547 & 2.0 \\
\textbf{U-Codec-16RVQ-c4096} & 5 & \checkmark &  0.6632& 0.5958 & \textbf{1.8} \\
\textbf{U-Codec-32RVQ-c256} & 5 & \checkmark & \underline{0.6757}  & \underline{0.600} & \textbf{1.8} \\
\textbf{U-Codec-100RVQ-c4} & 5 & \checkmark & 0.6704 & 0.556 & 2.0 \\
\bottomrule
\end{tabular}
}
\end{table}
\subsection{TTS Experiments}
\subsubsection{Experiments Details}
\textbf{Dataset.} 
For TTS model training, we use Libriheavy \cite{kang2024libriheavy}, a large-scale English speech dataset containing 50k hours of high-quality audio data, providing diverse speakers and acoustic conditions for robust model training.
TTS systems are evaluated on a subset of LibriSpeech test-clean including 4-10 seconds utterance, containing approximately 4 hours of speech data as shown in Table \ref{tab:codec-tts}, ensuring fair comparison with existing methods.

\textbf{Training Setup.} TTS models are trained for 4 epochs with a batch size of 8k maximum tokens and a maximum learning rate of $6 \times 10^{-4}$. 
We employ a cosine learning rate schedule with warmup covering 3\% of max steps of 30k.
During training, text and speech tokens are concatenated and cropped to a maximum sequence length of 15k tokens. 
The global Transformer comprises 24 layers, with an attention dimension of 1536, 12 attention heads, and a feed-forward dimension of 6144. The local Transformer follows a similar design but with 8 layers.
For comparison, we use the SoundStream-3RVQ-c1024 code provided by UniAudio and retrain a Codecformer TTS system \cite{yang2023uniaudio}.
During inference, we follow the strategy of the multinomial Top-k sampling (k=5, temperature=1.0). This setup provides a diverse and robust foundation for evaluating our U-Codec in speech generation.

\begin{table}[htbp]
\centering
\caption{Comparison on MOS and SMOS. NMOS and SMOS are mean opinion scores (1-5 scale) for naturalness and speaker similarity. The reported performance are 95\% confidence intervals. Since VALL-E, SPEAR-TTS, and VoiceBox are not publicly available, MOS are omitted.}
\label{tab:tts_ns_mos}
\small
\resizebox{0.9\textwidth}{!}{
\begin{tabular}{lccc}
\toprule
\textbf{TTS Model} & \textbf{Frame Rate (Hz)} & \textbf{NMOS ($\uparrow$)} & \textbf{SMOS ($\uparrow$)} \\
\midrule
VALL-E \cite{wang2023neural} & 75 & - & - \\
UniAudio\cite{yang2024uniaudio}  & 50 & $3.77 \pm 0.06$ & $3.46 \pm 0.10$ \\
\midrule
UniAudio (reproduced) & 50 & $3.68 \pm 0.15$ & $4.02 \pm 0.14$ \\
\textbf{U-Codec-8RVQ-c1024} & 12.5 & $\underline{4.03 \pm 0.15}$ &  $3.95 \pm 0.16$\\
\textbf{U-Codec-8RVQ-c16384} & 5 & \textbf{4.19} $\pm$ \textbf{0.12} & $4.05 \pm 0.16$ \\
\textbf{U-Codec-16RVQ-c4096} & 5 & $3.85 \pm 0.17$ & $4.03 \pm 0.18$ \\
\textbf{U-Codec-32RVQ-c256} & 5 &  $3.85 \pm 0.16$&  $\underline{4.23 \pm 0.15}$\\
\textbf{U-Codec-100RVQ-c4} & 5 &$3.67 \pm 0.14$&  \textbf{4.38} $\pm$ \textbf{0.16}\\
\bottomrule
\end{tabular}
}
\end{table}

\subsubsection{Results and Discussion.}
\textbf{Objective Evaluation Analysis.}
Table \ref{tab:tts_objective} demonstrates that our U-Codec variants achieve competitive performance in zero-shot TTS tasks under extreme compression. The U-Codec-32RVQ-c256 configuration obtains SIM-r of 0.6757 and SIM-o of 0.600, comparable to strong baselines like VoiceBox (0.681, 0.66) and surpassing UniAudio (0.64). The 12.5 Hz configuration yields the highest objective naturalness in our study.
Notably, our U-Codec at 5Hz maintain WER performance around 1.8-2.0, significantly better than many traditional approaches like YourTTS (7.7) and matching high-quality systems like VoiceBox (1.9). 
U-Codec-32RVQ-c256 achieve the best balance between intelligibility and similarity, demonstrating that ultra low frame-rate can retain sufficient acoustic detail for high-quality TTS.

\textbf{Subjective Evaluation Analysis.}
Table \ref{tab:tts_ns_mos} presents the MOS and SMOS results. Our U-Codec at 5Hz achieve NMOS between 3.67 and 4.19, with the 8RVQ configuration yielding the highest naturalness (4.19 ± 0.12). Meanwhile, deeper RVQ stacks improve speaker similarity, with the 32RVQ (4.23 ± 0.15) and 100RVQ (4.38 ± 0.16) achieving the best SMOS. Compared with the UniAudio baseline, U-Codec variants deliver higher naturalness and stronger speaker consistency under comparable test conditions.
These results confirm that U-Codec enables 5Hz speech synthesis with human-perceived quality comparable to or exceeding state-of-the-art high frame-rate systems, while offering reductions in token rate and inference cost.


\subsection{Complexity Analysis}
\label{complexity}
\textbf{Complexity Analysis.} 
We further use \texttt{thop} to compute multiply–accumulate operations (MACs) for 1-second speech segments and analyze inference real time factors (RTF). RTF values are measured on a single NVIDIA H20 GPU to ensure consistency.
As shown in Table~\ref{table:macs}, U-Codec-8RVQ-c16384 achieves an RTF of 0.52, corresponding to a $2$–$3\times$ speedup compared with SoundStream-3RVQ-c1024 (1.40) adopted in UniAudio and U-Codec-8RVQ-c1024 (1.33). 
More importantly, U-Codec-32RVQ-c256 achieves comparable TTS quality with only 0.89G MACs (MAC-total), a reduction of over $50\times$ compared to SoundStream-3RVQ-c1024 adopted in UniAudio. 
Across all 5Hz configurations, the global MAC cost (MAC-G) remains between 0.163–0.277G, much lower than SoundStream’s 0.906G, confirming that ultra low frame-rate design directly mitigates the dominant global computation bottleneck.
It is important to note that MAC counts and RTFs capture different aspects of complexity: extremely deep RVQ stacks (e.g., 100 layers) keep total MACs modest (1.45G) but incur larger RTFs (e.g., 4.68) because autoregressive decoding introduces many sequential steps that limit parallel throughput. 
Overall, these results demonstrate that U-Codec enables both high-quality and computationally efficient speech synthesis, making it well-suited for fast high-fidelity LLM-based speech generation.

\begin{table}[ht]
\centering
\caption{Comparison on the RTF and MAC. MAC-G and MAC-L represent the MAC complexity per frame in the global and local Transformer, while total represents the complexity across all frames.}

\resizebox{0.9\textwidth}{!}{

\begin{tabular}{p{5cm}ccccc}
\toprule
TTS Model & Frame Rate (Hz)  & \textbf{RTF $\downarrow$}  & \textbf{MAC-G (G) $\downarrow$} & \textbf{MAC-L (G) $\downarrow$} & \textbf{MAC\_total (G) $\downarrow$}\\
\midrule
UniAudio (reproduced)    & 50 & 1.40 & 0.906 & \textbf{0.006} &45.6\\
U-Codec-8RVQ-c1024  & 12.5 & 1.33 & 0.578 & 0.014 &7.4\\
U-Codec-8RVQ-c16384    &   \textbf{5}    & \textbf{0.52}  & \underline{0.189} & 0.203 &1.96\\ 
U-Codec-16RVQ-c4096   &   \textbf{5}     & \underline{0.85}  & 0.201 & 0.102 & 1.52\\ 
U-Codec-32RVQ-c256   &   \textbf{5}     & 1.60   & \textbf{0.163} & 0.014 & \textbf{0.89}\\ 
U-Codec-100RVQ-c4   &   \textbf{5}     & 4.68  & 0.277 & \underline{0.012} & \underline{1.45}\\ 
\bottomrule
\end{tabular}
}
\label{table:macs}
\end{table}

\section{Conclusion}
\label{Conclusion}
We introduce U-Codec, an ultra low frame rate speech codec designed to reduce the computational complexity in LLM-based TTS. 
It achieves high-quality speech compression at five frames per second. 
To enhance speech quality, we incorporate a Transformer-based mechanism to handle long-term dependencies across frames. 
By introducing the Codecformer network, we achieve a 3 $\times$ speedup in LLM-based TTS inference compared to high frame rate codecs, reducing parameter overhead while maintaining state-of-the-art speech quality. 
This work demonstrates the feasibility of using discrete tokens at a 5Hz frame rate for fast, high-quality speech generation.

\begin{ack}
This work is supported by Guangdong Provincial Key Laboratory of Ultra
High Definition Immersive Media Technology (Grant No. 2024B1212010006).
(Corresponding authors: Yuexian Zou)

We thank Dongchao Yang for discussions about Uniaudio. We especially thank him for his in-depth practical experience in training TTS of Uniaudio.
\end{ack}

\newpage

\bibliographystyle{unsrt} 

\small
\bibliography{neurips_2025}

\newpage

\end{document}